# Follow the user meaningfully and product growth will follow
A mixed methods case study tying UX Point of View & Growth leading to measurable impact

Neha Raghuvanshi
neha.n.raghuvanshi@gmail.com

## ABSTRACT

Have you wondered how cross-functional teams balance between maximizing value that users derive and business growth leading to win-win situations? This case study shows how User Experience Research (UXR) and Data Science teams used mixed methods research to strategically influence Product Led Growth (PLG) for a Password Manager used by million+ users, thus allowing our users, internal teams, and business to win. The audience will take away practical lessons/techniques related to leveraging mixed methods to:
a. Maximize user value while meeting business growth goals
b. Influence cross-functional teams
c. Measure user and business impact

This case study can be easily tied to the UXR Point of view pyramid (POV) [2] that represents a methodological approach to construct a POV and further dives into actioning POV to create measurable user and business impact.

1.  **INTRODUCTION**
    Product-led growth (PLG) is a business methodology in which user acquisition, expansion, conversion, and retention are all driven primarily by the product itself. It creates company-wide alignment across cross-functional teams around the product [5]. As per PLG framework [5], the product experience is designed to generate higher user satisfaction and increased advocacy, which in turn drives compounding growth of new user acquisition.

    A newly formed team including Product, UX, Data, Marketing was tasked with accelerating user growth and revenue in a classic scenario where "Who a product was built for" (IT Admin) is slightly different from "Who is using it" (Everyone).

This case study highlights how UXR and Data Science partnered to advocate for 'who we build for' in a numbers-centric organization. The UXR POV championed for an underserved group of customers aka "lean back" [1] whose needs weren't being met by more technically driven messaging and feature sets that typically catered to technical and security conscious users, aka "lean in". Discovering the right audience impacted the Targeting, Messaging, & Product strategy ultimately leading to measurable growth.

This case study details how one foundational user research study focuses on the research goal "Who is using What and Why?" led to a unique and holistic POV, further opening up multiple research opportunities to action and solidify research POV leading to measurable user, business & cross-functional impact.

The user and business impact or growth goal can be thought of as a dish a chef wants to prepare. The key steps followed to run & create impact using Foundational Research can be considered as the secret sauce used to build a recipe.

The multiple research streams that opened up can be thought of as tried and tested recipes that used different mixed methods to drive user, business, & cross-functional team impact.

Before diving into the recipes, it makes sense the 6-step secret sauce across the user research lifecycle leading to Product Led Growth. This secret sauce nicely ties to the phases of UXR PoV Pyramid - Foundational, Data Collection, Insight Generation, & UXR POV [2].

## 1.1 SECRET SAUCE

| Secret Sauce | Best Practice | UXR POV Pyramid Phase [2] |
|---|---|---|
| **1. Seek BUY IN for Foundational Research using Usage Data (telemetry)** | Product usage data (telemetry) is a great way to get stakeholder buy in to propose user research studies and show impact post study. It also helps with targeted recruiting. | Foundational |
| **2. Create EMPATHY & ENGAGEMENT using Qualitative Research** | Foundational interviews often take time to wrap up and it is not possible to include all the stakeholders during data analysis, hence, sharing soon can not only help generate empathy among teams who don't talk to users often but also leads to stakeholder involvement in co-creating insights or uncovering follow-up questions for future sessions. | Data Collection |
| **3. ANALYSE data by tying Qualitative & Usage data to business goals** | Thinking of what user data observations or insights mean to your business can help you develop unique perspectives such as identifying distinct user groups particularly in products that serve one common need or job to be done e.g. storing and using passwords where finding distinct groups of users is not straightforward. | Insight Generation |
| **4. Build TRUST & CONFIDENCE using Quantitative Data** | Leveraging quantitative tools such as surveys can be powerful in scenarios where qualitative insights are making sense but are not enough to make pivotal strategic actions. Running something quick and small can help solidify qualitative insights. | UXR POV that expedites impact |
| **5. INSPIRE actionable Ideas by tying Qualitative Data to Growth/Business Goals** | Thinking of research insights or UXR POV in terms of business goals as a starting point may hinder creative thinking. Thinking purely from a users' lens may be inspiring but may not necessarily land with business. Hence, leverage research insights for creativity but do not view them in silos only. Demonstrate how they tie to stakeholder specific goals e.g., Conversion, Reliability by leveraging existing projects & rituals will help them land and generate a stronger and holistic Point of View. | UXR POV that communicates effectively |
| **6. Take ACTION and measure Impact using Cross-functional collaboration, Quantitative & Usage Data** | 1. Amplify research insights during cross-functional rituals e.g., Engineering stand ups, Design sprints, Marketing campaign planning<br>2. Work closely with Product, Engineering, Marketing, Data, & Design teams to measure impact due to UXR by:<br>a. Running experimentation e.g. A/B tests<br>b. Instrumenting in-product surveys that measure user experience & attitudes<br>c. Simultaneous triangulation between behavioral data (telemetry) and attitudinal data (in product surveys) can help paint a holistic picture and longitudinally measure progress of user experience and cross-functional teams.<br>d. Leveraging survey verbatim themes to measure impact | UXR POV that drives impact by validation through cross-functional collaboration |

## 1.2. Mixed Methods UXR Recipes to potentially measure & impact UXR POV leading to growth

| DISH | | RECIPES | | | |
|---|---|---|---|---|---|
| | Foundation | | Data Collection & Insight Generation | | POV |
| Growth Goal | Research Outcome | Telemetry | Qual | Quant | |
| 1.Acquisition | Identify Target Audience | Cohort Data | Personas | Persona Distribution Survey | Consider underserved segment |
| 2.Activation | Identify User-Centric Messaging Strategy | A/B Tests | Insights Framework | | Leverage collaborative moments to show value via messages |
| 3.Adoption | Improve Reliability by prioritizing engineering fixes | Usage Triggers | Survey Verbatim | In Product Reliability Survey | Minimize friction due to unreliability |
| 4.Revenue* | Define Product Vision | | Insights Framework | Opportunity Sizing Survey | Consider new partnerships |
| 5.Advocacy* | Formulate Onboarding Strategy | | Insights Framework + Usability Tests | Usability + Effort Score | Include hand holding in Experiences |

*Potential measurements based on opportunity sizing or engineering data prediction

## 1.3 Examples of measuring potential user, business & cross-functional team impact based on UXR POV

| Cross Functional Teams | Marketing | Design | Engineering | Leadership |
|---|---|---|---|---|
| UXR Recommendation (POV) | Leverage collaborative moments to show value via messages | Include hand holding in experiences | Minimize friction due to browsers | Consider new partnerships |
| Experience Metrics | Feature use | Effort Score | Reliability* | Quantified Unmet needs* |
| Business Metrics | Virality/Acquisition | Activation | Adoption* | Revenue* |

*Potential measurements based on opportunity sizing or engineering data prediction

## 2. 6-STEP SECRET SAUCE

### 2.1. Seek Buy In for Foundational Research using Usage Data (telemetry)
(UXR POV Pyramid Phase - Foundation)

In feature-centric teams, getting buy-in for foundational research is often difficult as teams find it hard to imagine tangible outputs, however the idea of giving meaning to our behavioral clusters based on our user activity in our product (telemetry) got our Data and Leadership teams excited. This worked out perfectly from a research standpoint because it is often hard to figure out where to start recruiting in a large base of million+ users. Taking advantage of the behavioral clusters based on product usage allowed us to expedite diverse and representative recruitment.

We aligned on the research goal to understand "Who is using the product actively, What are they using and Why?"

*Best Practice: Product usage data is a great way to get stakeholder buy in to propose user research studies and show impact post study. It also helps with targeted recruiting.*

## 2.2. Create Empathy & Engagement using Qualitative Research
(UXR POV Pyramid Phase - Data Collection)

We ran around ~30 qualitative interviews and it wasn't possible for each team member to attend. However, post-session debriefs with stakeholders were helpful in thinking of the observations from the lens of stakeholders and business, ultimately helping during the insight generation phase.

Moreover, sharing post session takeaways as a narrative related to each participant via chat led to a lot of engagement from our cross-functional teams and leadership. While some members of our teams started associating their password storing and sharing habits with individual participants e.g. an engineer said his habits were like Participant 1 and he could totally relate to him, other team members followed up with some great follow up questions, which were helpful in the insight generation phase.

*Best Practice: Foundational interviews often take time to wrap up and it is not possible to include all the stakeholders during data analysis, hence, sharing soon can not only help generate empathy among teams who don't talk to users often but also leads to stakeholder involvement in co-creating insights or uncovering follow-up questions for future sessions.*

## 2.3. Analyze data by tying Qualitative & Usage data to growth/business goals
(UXR POV Pyramid Phase - Insight Generation)

30 interviews can produce a lot of data and there may be many interesting nuggets there. As researchers, we can often get overly passionate in championing for the user's voice, however, our takeaways may not always land with stakeholders if we don't tie it to their key objectives. Hence, thinking of research observations or insights in terms of what it 'means' to our business helped us generate unique insights and POV.

### 2.3.1 Using meaningful analysis to synthesize personas

Our first goal was to identify distinct user groups. To move fast, we imposed pre-set theories or codes on the data to make it easier to synthesize personas e.g., we tagged observations as needs, pains, goals, motivations hoping to find patterns between user groups. We soon realized that the most 'interesting' codes/tags based on insights/themes generated from this research, typically known as inductive analysis, were not captured within the personas when we used shortcuts to analyze personas. This was primarily because for a product like a password manager, the needs, goals, pains, and motivations are mostly the same across user groups i.e., everyone wants to browse the internet securely, conveniently, and easily. Once we were able to, think of research observations or insights in terms of what it 'means' to our business in terms of user acquisition, feature activation or adoption, product purchase, upgrade, or referral enabled 'meaningful' data analysis. Our inductive themes showed us the most 'meaningful' differences are due to:

a. Security and technology-related behaviors and attitudes
b. How do users enable or refer others
c. Situations in which they use, buy, upgrade

These themes allowed us to represent our qualitative personas using 8 actionable dimensions, enabling us to distinguish between our distinct user groups by rating them across scales related to these dimensions. An example of a dimension is 'help users get started with the product' and a related scale would be rating users' groups on how often they help users getting started with the product – Always, Sometimes, Never. These meaningful differences were abstracted as memorable persona frameworks - Lean in and Lean Back users. While our personas had details about our users' goals, needs, pains etc., looking at the comparison of the 4 personas across the 8 dimensions was a great way for our teams to get a snapshot of our users.

### 2.3.2. Using Meaningful Analysis to shape referral and onboarding experiences

**Observation: What did we learn from interviews?**
Some Lean in users personally guide and nudge Lean Back users to get them started, others act on the Lean Back user's behalf

**Meaningful synthesis: What does it mean to our business?**
Recognizing users who enable others by guiding vs. acting on their behalf will allow Product and Marketing to shape referral programs. Users who personally guide others may be the strongest advocates for our product.

This can further allow product and design to translate in-person handholding experience of users that show to in-product self-served experiences, thus increasing product and feature activation/adoption.

Thus, adapting to both Users and Business Needs by pivoting to 'meaningful' research analysis helped us nail "Who we build for" and hone our research rigor.

While the Lean in and Lean back persona framework was etched into the minds of our stakeholders, teams were not sure what the distribution of Lean in and Lean back users looked like, they wondered if it was safe to change our Target Audience based on 30 interviews.

*Best Practice: Thinking of what user data observations or insights mean to your business can help you develop unique perspectives such as identifying distinct user groups particularly in products that serve one common need or job to be done e.g. storing and using passwords where finding distinct groups of users is not straightforward.*

### 2.4. Build Trust & Confidence using Quantitative Data
(UXR POV Pyramid Phase - UXR POV that expedites impact)

We could expedite application of personas across the product once we validated "lean in" and "lean back" qualitative personas created based on 30 in-depth interviews by running a quantitative cluster analysis on 2500 survey responses. Once stakeholders saw the distribution of personas across the product, they recognized what we are losing out by not catering to "lean back".

Based on the quantitative work we could also narrow down the 8 dimensions to 4 and recognize the ones that are most important to show differences between our personas. Narrowing down dimensions is particularly helpful when recruiting future research participants, targeting specific users during and personalizing experiences.

Thinking about personalization, imagine users fill out an in-product quiz based on these 4 dimensions, and we identify the strongest advocates of our product. We not only provide these advocates with personalized experiences where they can enable others, but we can also find product usage patterns among advocates, eventually recognizing what product functionality matters the most to them. This would allow us to systematically tie behaviors and attitudes, giving us a holistic view of our users in the product, and further product led growth.

Thus, adapting to our stakeholders' needs by the timely triangulation of data expedited changes in targeting strategy (Who we build for?), experience strategy (What we build?), and messaging strategy (How we communicate?).

While the quantitative piece helped us associate magnitudes or leverage most important criteria, qualitative pieces inspired future-forward ideas that fed into this strategy.

*Best Practice: Quantitative tools such as surveys can be powerful in scenarios where qualitative insights are making sense but are not enough to make pivotal strategic actions. Running something quick and small can help solidify qualitative insights* and can expedite action.

### 2.5. Inspire actionable Ideas by tying Qualitative Data to Growth/Business Goals
(UXR POV Pyramid Phase - UXR POV that communicates effectively)

We wanted teams to stay inspired by these personas and didn't want them to eventually end up as a Share Point document that is not frequently visited. "Who are our users" was etched in the memories of our teams, but how do they tie this to tangible examples of "How they use the product" was something that felt like a missing piece of the puzzle. The rich insights from the in-depth interviews were a wonderful way to highlight how personas come into play.

We first considered clustering 40 insights from in-depth interviews based on business metrics or features, so we talked the stakeholders' language, however, the most interesting qualitative insights were buried deep and would take away focus from the users if we led with business/features first. This led to the inception of a memorable qualitative insights' framework, P-A-G-E, that was based on organizing the 40 insights around the most interesting user insights. The framework personifies the value of [App/Service/Company] as Planner, Aide, Guide, and Ensurer, going beyond the typical value-drivers of providing Security or Convenience, that are prevalent in the cyber-security space. Stakeholders could go over any of these tenets in a 'Choose your Own Adventure' (CYOA) style [4] and thus think of Password Management from different angles.

#### 2.5.1. User Insights Framework example

Here is an example of how the PAGE tenet Planner came about: we observed exchange of credentials increased among couples/families during collaborative life events such as weddings, childbirth, home-buying etc. or emotional moments such as an aging family member's healthcare. Thus, thinking of the password manager as a Planner, would allow users to plan for collaborative or emotional moments in life, by having the credentials of loved ones handy and 'share' them.

#### 2.5.2. Effectively communicating UXR POV

This strategic insights framework allowed our teams to connect the Who (persona), What (features/needs), Why

(value and led to stakeholder engagement, but there was always a question 'so what'. Tying this framework to Growth Goals, Features & Personas using the Choose your own Adventure (CYOA) [4] led to a holistic Point of View.

For example, exploring the adventure of the password manager as a 'Planner', Lean In users could leverage collaborative life events to convince Lean Back users to activate the product or specific features (e.g., sharing using this product. Features like Sharing allowed one user to introduce the product to other users as they had to accept credentials such as Netflix passwords, banking passwords etc. through the product, thus increasing user acquisition, ultimately leading to Product Led Growth.

Thus, adapting our presentation approach, by keeping our users at the heart of it, and positioning our work in a way stakeholders see instant benefit helped us keep cross-functional teams engaged and inspired. We led with the user's voice, layered in recommendations related to business metrics and feature-related details, ultimately enabling a shared language based on P-A-G-E.

*Best Practice: Thinking of research insights or point of view in terms of business goals as a starting point may hinder creative thinking. Thinking purely from a users' lens may be inspiring but may not necessarily land with business. Hence, leverage research insights for creativity but do not view them in silos only, ultimately demonstrating how they tie to stakeholder specific goals e.g. Conversion, Reliability will help them land and generate a stronger and holistic Point of View.*

## 2.6. Take action & measure impact using cross-functional collaboration, quantitative & telemetric data

(UXR POV Pyramid Phase - UXR POV that drives impact by validation through cross-functional collaboration)

Here are a few examples or recipes of how we helped teams co-create POV through cross-functional collaboration and measured impact systematically.

### 2.6.1. Recipe 1: Marketing to increase feature activation and user acquisition

Our Marketing team had to plan campaigns for the holidays, the UXR team helped them set up ideation workshops for messaging ideas to inspire ideas among cross-functional teams based on needs of Lean in and Lean Back Personas, PAGE Insights framework & Growth Goals. Here is an example of how UXR & Marketing co-created POV during these workshops that increased feature activation & user acquisition.

#### 2.6.1.1. Example Insight
One way to acquire new users into the product is by introducing them to Sharing. Based on the research we could identify what's stopping users from activating, (i) either they are unaware about the usefulness of sharing, (ii) or they think sharing passwords via texts or shouting them aloud needs less effort. We also learnt users are motivated to have passwords handy, so they are not locked out of their favorite online accounts when someone changes a password e.g., Netflix.

#### 2.6.1.2. Co-creating UXR POV
We recognized that if we trigger a Marketing campaign to highlight the usefulness of Sharing via the Password Manager and how it's better than texting passwords at the right moment, we can encourage users to share.

#### 2.6.1.3. Actioning UXR POV

Marketing with inputs from UX Design, Writing, and Product, thus leveraged the idea of Password Manager as a Planner during holiday seasons and highlighted the idea of sharing streaming and shopping passwords safely among family members to be better planned for the holidays. Having a shared language based on a user's voice, generated synergy and creativity within teams, thus allowing them to win.

#### 2.6.1.4. Measuring UXR POV Impact

On A/B testing with the original feature-centric campaign that focused on **"how sharing works"**, this research inspired campaign highlighted **"why to share"**, leading to increase in content engagement (increase in email open rate), Sharing feature activation (increase in users who shared) thus leading to an increase in user advocacy and acquisition (increase in users that downloaded the password manager to accept the shared item), ultimately leading to potential business growth. Increase in the use of sharing based on the marketing campaign implies users now see value in the feature and use it, hence allowing the product to provide value to users and helping them win.

### 2.6.2. Recipe 2: Engineering to increase product Reliability

Ideation workshops were a means to inspire Marketing, getting involved in Engineering sprint planning activities was a means to casually slip in the idea of thinking of a Password Manager as an Aide while browsing the internet.

The original feature-centric engineering team focused on just one part of the product, and could now easily see how different parts of the product now connected when they started thinking of a Password Manager as an Aide.

#### 2.6.2.1 UXR POV Pyramid - Foundation
The engineering and UXR team aligned on the idea that for an Aide to be successful, it needs to be reliable, this led to a new UXR workstream to enhance product reliability.

### 2.6.2.2. UXR POV Pyramid - Data Collection
Based on our usage data, Data & Engineering teams had hypotheses of scenarios the product is not working reliably. Hence, we were able to trigger in-product surveys during those scenarios and not only used it to validate our hypothesis but were also able to make incremental improvements to improve the reliability of our product.

### 2.6.2.3. Insights and Point of View
A combination of the in-product survey that measured experiences quantitatively with open ended responses that helped us hone into experiences qualitatively ultimately helped us measure progress on improving reliability longitudinally. We could also use the research data to validate & challenge some engineering hypotheses based on usage data, hence they tweaked how we collected some telemetric data.

The foundational project with a holistic Point of View led to actionable insights and future streams of research, ultimately enabling us to measure potential impact from a user's perspective & business (product growth). We've only covered two detailed examples (recipes) of measuring impact based on user research in this paper. This table summarizes how we would potentially measure cross-functional impact, user experience & business growth.

| Xfn Teams | Marketing | Engineers | Design | Leaders |
|---|---|---|---|---|
| **UXR Recommend-ation Examples** | Leverage collaborative moments to show value via messages | Minimize friction due to unreliability | Include hand holding in product experience | Consider new partnerships |
| **Experience Metrics** | Feature use | Reliability | Effort Score | Quantified Unmet needs |
| **Business Metrics** | Virality | Adoption | Activation | Revenue |

*Potential measurements based on opportunity sizing or engineering data prediction

*Best Practice:* 1. Amplify research insights during cross-functional rituals E.g., Engineering stand ups, Design sprints, Marketing campaign planning etc. 2. Work closely with Product, Engineering, Marketing, Data, & Design teams to measure impact due to UXR by:
a. Run experimentation e.g. A/B tests
b. Instrument in-product surveys that measure user experience & attitudes

c. Simultaneous triangulation between behavioral data (telemetry) and attitudinal data (in product surveys) can help paint a holistic picture and longitudinally measure progress of user experience and cross-functional teams leveraged survey verbatim themes to measure impact

## CONCLUSION & FUTURE WORK
Product, Design, Marketing, and Engineering applied UXR POV inspired from the Lean in-Lean Back personas and P-A-G-E insights framework and could impact other user-centric metrics e.g., Usefulness, Effort it takes to use this Password Manager, Reliability etc. and business metrics such as Acquisition, Activation, Adoption, Revenue, and Referral ultimately leading to PLG where users, teams, and business wins. The secret sauce here was inspiring with Qualitative Research, expediting research application with Data and Quantitative Research. The audience will take away practical lessons and best practices related to mixed methods, designing for growth and engagement we have used during the various stages to balance between user-centricity and business centricity.
We've only elaborated on two examples (recipes) related to measuring impact for brevity, we can get into details for other recipes.